\documentclass[a4paper,12pt]{article}

\usepackage{amsmath,amssymb}
\usepackage{graphicx}

\newcommand{\dd}{{\mathrm d}}
\newcommand{\ceff}{{\mathcal{E}}}

\begin{document}

\title{Nonlocal interactions in coagulating particle systems}
\author{T.H.M. Stein and S.V. Nazarenko\\Mathematics Institute, University of Warwick\\Coventry CV4 7AL, United Kingdom\\t.stein@warwick.ac.uk}

\maketitle

\section*{Abstract}

We consider a three dimensional system consisting of a large number
of small spherical particles, which move due to gravity or with laminar shear and which merge when they cross. A size ratio criterion may be applied to restrict merging to similar sized particles (locality of interactions) or particles dissimilar in size (nonlocality). We perform direct numerical simulations (DNS) of this particle system and study the resulting mass spectra. In mean field approximation, these systems can be described by the Smoluchowski coagulation equation (SCE). DNS of the particle system with locality enforced show the scaling solutions or Kolmogorov-Zakharov spectra for the SCE, signifying a constant mass flux. DNS without a size ratio criterion show $-4/3$ scaling for large particles in a system with gravity, signifying a constant flux in number of particles, which we also find analytically by assuming nonlocality of interactions in the SCE. For laminar shear, this nonlocality is only marginal, and our DNS show that a correction to the scaling solution is required. 


\section{Introduction}
\label{sec:intro}

Systems of coagulation are abundant in the physical sciences, with notable examples in the atmosphere~\cite{fried1,westbrook}, and applications in chemistry~\cite{hunt,weitz} and astrophysics~\cite{vkon}. Our initial interest in these systems was raised by their use in the theory of warm rain formation~\cite{falk,jeffrey,pruppacher} and the existence of scaling solutions~\cite{hunt,junge1}. 

We consider coagulation as an event where two particles of mass $m_1$ and $m_2$ merge to form a particle of mass $m=m_1 + m_2$. The exact nature of how these particles meet and merge is described by a collision kernel $K(m_1,m_2)$. This description is reminiscent of the turbulence formalism, where wave interaction between wave numbers $k_1$ and $k_2$ is described by an interaction coefficient $V_{k12}$~\cite{ZLF}. We wish to continue this analogy, and create a particle system with a turbulence setup, i.e. with particles forced at some small mass $m_0$, and removed once they reach a large mass sink $M$. Assuming this system reaches a steady state, we hope to find scaling solutions conform Kolmogorov-Zakharov solutions in turbulence.

From a mathematical perspective, particle coagulation has been discussed with a focus on scaling and self-preserving solutions to the Smoluchowski coagulation equation (SCE), which was originally derived as a mean field approximation to the physical coagulation processes for Brownian motion and laminar shear~\cite{smol}. Scaling solutions for general kernels of coagulation have been classified, with a particular interest for unphysical solutions that imply gelation, i.e. the formation of a particle of infinite mass at finite time~\cite{dongen,dong1,dong2}. Similarities between the SCE and turbulence theory have been discussed previously~\cite{pushkin}, with an explicit discussion of locality of interactions. In this paper, we use results from an interacting particle system to highlight the importance of this assumption of locality when deriving scaling solutions. 

First, in Sect.~\ref{sec:ips} we will introduce the interacting particle system, which does not assume mean field behaviour or any other properties of the SCE. The system can model particles moving due to gravity, a coagulation process known as differential sedimentation~\cite{hunt}, or particles moving due to laminar shear. Using a simple size ratio criterion, we can ensure only similar-sized particles merge, making sure the locality assumption is met, or we ignore the size ratio and allow all particle interactions. We also perform simulations in which we only allow interactions between particles very dissimilar in size to study nonlocal interactions more directly.

We continue by introducing the SCE and provide the well-known scaling results (Sect.~\ref{sec:mf}). We comment on further properties of the kernels for laminar shear and differential sedimentation, in particular the convergence of the collision integral in the SCE (Sect.~\ref{sec:convergence}).

In Sect.~\ref{sec:ds} we revisit the approximate solution for the system with dominance of nonlocal interactions in differential sedimentation~\cite{horvai}. Using the IPS, we retrieve the scaling solution. We find that the system without forced locality has a bottleneck and that it shows non-mean field behaviour. We provide several techniques to reduce this finite size effect and compare results to the nonlocal solution.

Finally, we study laminar shear in Sect.~\ref{sec:ls}, which in terms of the SCE is known to be a boundary between gelling and non-gelling systems. We do not provide a nonlocal solution for this case, but we present numerics from the IPS showing a deviation from the scaling solution that clearly is a result of nonlocal interactions.


\section{The interacting particle system}
\label{sec:ips}

We have developed a very simple system of a large number of interacting particles (IPS) with which we can easily test a variety of properties of coagulation. We use a setup typical for turbulence simulations, in which we force mass at small scales and dissipate at large scales, and let the mass or volume spectrum develop.~\footnote{We will make no formal distinction between mass and volume in this paper.} Mass transfer occurs because of particle coagulation, where two particles of volumes $m_1$ and $m_2$ form a new particle of volume $m=m_1+m_2$. We may allow or disallow a coagulation event between two particles depending on their size ratio $m_1/m_2$, thus enforcing either locality or nonlocality of interactions. If we allow all particle interactions, we refer to this system as ``free merging''.

In the IPS, each time step we introduce a fixed number $N_0$ of particles of discrete multiples $j$ of the minimum particle volume $m_0$ with $j \in [1,5]$, uniformly distributed in space. This space is a periodic box of size $L_x \times L_y \times L_z$, where typically $L_x=L_y=L_z=10\mathrm{cm}$. We consider the particles spherical with minimum radius $r_0 = 0.01\mathrm{cm}$ and to have no interaction with the surrounding ``fluid''. Particle coagulation in the IPS happens when the distance between the centers of two particles of volume $m_1$ and $m_2$ is smaller than the sum of their radii $r=r_1 + r_2$. A new particle of volume $m=m_1+m_2$ is then created at the center of mass of the original particles. If the rules of collision disallow this event, the particles will cross one another, having no effect on each other's trajectory. Finally, we typically choose a strict cutoff $M=10^4 m_0$, so that any particles formed that have $m>M$ are removed from the system at the end of the time step.

In a system with gravity, large particles will overtake smaller ones because of their higher terminal velocity, and thus coagulation may occur. This system is referred to as differential sedimentation~\cite{jeffrey}. We assume particles move at Stokes terminal velocity $u\sim G r^2 \sim m^{2/3}$, where we choose $G=2\cdot 10^3 \mathrm{cm}^{-1}\mathrm{s}^{-1}$, so that a particle of radius $0.1\mathrm{cm}$ travels at $20\mathrm{cm}\cdot\mathrm{s}^{-1}$, equivalent to an air bubble rising in water~\cite{Clift}. 

For laminar shear, we choose a shear function $f(z)=\gamma z$ with $\gamma=12.5 \mathrm{s}^{-1}$, so that those particles at $z=10$ travel at $125 \mathrm{cm}\mathrm{s}^{-1}$. Particles at $z'=z+\dd z$ will overtake those at $z$, and coagulation may occur. To ensure periodicity between particles near $z=0$ and those near $z=L_z$, when we check for these events we move all particles with $0 < z' < 0.5$ to $z' + L_z$. The $x$-position of those particles is then altered so that $x_{new} = x_{old} + \gamma T L_z$, where $T$ is the time that has elapsed since the start of the simulation. Once we have checked for collisions over this periodic boundary, we move the particles back to $z'$ and transform $x_{new}$ back to $x_{old}$.

Simulations for a system with free merging will have no collision rule applied. In systems with forced locality or nonlocality, we check the size ratio $m_1/m_2$ between two overlapping particles. If this satisfies $1/q < m_1/m_2 < q$, where typically $q=2$, we have locality. If we only allow particle coagulation when the above criterion is satisfied, we call this a system with ``forced locality''. If we only allow particle coagulation when the above criterion is not satisfied, we call this a system with ``forced nonlocality''.

At long times, we assume the IPS has reached a statistical steady state, and we produce the particle size spectrum. We divide the volume range into logarithmic bins $b_k$ with $k\in \mathbb{N}$ such that $b_k=\left[1.1^k m_0,\right. \left. 1.1^{k+1} m_0 \right)$. At each time step, we add all particles to their corresponding bin $b_k$ and after $10,000$ time steps we average to find the particle size spectrum. This method smooths the resulting spectrum and has no noticeable effect on the scaling.


\section{Mean field approximation}
\label{sec:mf}

The Smoluchowski coagulation equation~\cite{smol} has been applied to numerous aggregation and clustering mechanisms in physics and chemistry~\cite{fried1,hunt,mccave,vkon}. We will check whether it is an appropriate mean field approximation for the IPS. We write the SCE
\begin{align}
\label{eq:smol}
	\frac{\partial n}{\partial t} + \textbf{u} \cdot \nabla n
	= 
	& \frac{1}{2} \int_0^{m} \hspace{-3mm} \dd m_1 K(m_1,m-m_1) n(m_1) n(m-m_1)
\\
\nonumber
&
- \int_0^{\infty} \hspace{-3mm} \dd m_1 K(m,m_1) n(m) n(m_1) 
\ ,
\end{align}
where $n(m)$ is the density of particles of volume $m$. The function $K(m_1,m_2)$ is the collision kernel for particles $m_1$ and $m_2$, which describes the collision process between particles merging to form a particle of size $m_1 + m_2$, and $\textbf{u}$ is the particle velocity. 

In past research, scaling solutions to the SCE have been derived using dimensional analysis~\cite{hunt,jeffrey}. These solutions have also been derived by application of the Zakharov transformation to the collision integral~\cite{zakfil}, and conform turbulence terminology they can be interpreted as a Kolmogorov-Zakharov cascade of volume or mass~\cite{crzab,vkon}. For differential sedimentation this solution is
\begin{equation}
\label{eq:ds}
	n(m)
\sim
	m^{-13/6}
\ ,
\end{equation}
and for laminar shear 
\begin{equation}
\label{eq:shear}
	n(m)
\sim
	m^{-2}
\ .
\end{equation}
Both the scaling argument and the Zakharov transformation rely on locality of interactions, i.e. convergence of the collision integral in the SCE. However, we will now show that both the kernel for differential sedimentation and the one for laminar shear have scaling exponents that lead to divergence.


\section{Convergence of the collision integral}
\label{sec:convergence}

Let us rewrite the SCE~\eqref{eq:smol} for a system with a small and large mass cutoff, i.e. a truncated spectrum~\cite{klett}
\begin{align}
\label{eq:cuts}
	\frac{\partial n}{\partial t} + \textbf{u} \cdot \nabla n
= 
	 &\int_{m_0}^{m/2} \hspace{-3mm} \dd m_1 K(m_1,m-m_1) n(m_1) n(m-m_1)
\\
\nonumber
& - \int_{m_0}^{M} \hspace{-3mm} \dd m_1 K(m,m_1) n(m) n(m_1) 
\ ,
\end{align}
where we have used the symmetry of the first integral.

We assume a scaling solution to this equation of the form $n(m)\sim m^{-x}$, which requires locality of interactions, or collision of the integral as $m_0 
\downarrow 0$ and $M\rightarrow \infty$. Also, we assume two types of scaling behaviour of the kernel, namely
\begin{equation}
\label{eq:scaleK}
	K(m x_1, m x_2)
=
	m^{\lambda} K(x_1,x_2)
\ ,
\end{equation}
and for $x_1 \gg x_2$
\begin{equation}
\label{eq:scaleKlarge}
	K(x_1,x_2)
\sim
	x_1^{\lambda} x_2^0
\ .
\end{equation}
Let us introduce $m_-$ and $m_+$ such that
\begin{equation}
\label{eq:inequals}
m_0 \ll m_- \ll m \ll m_+ \ll M
\ ,
\end{equation}
and we rewrite eq.~\eqref{eq:cuts} using these new delimiters
\begin{align}
\label{eq:rewrite}
	\frac{D n}{D t}
=
	& \int_{m_0}^{m_-} \mathcal{K}(m_1,m-m_1) - \mathcal{K}(m_1,m) \dd m_1
\\
\nonumber
	& + \int_{m_-}^{m/2} \mathcal{K}(m_1,m-m_1) - \mathcal{K}(m_1,m) \dd m_1
\\
\nonumber
	& - \int_{m/2}^{m_+} \mathcal{K}(m_1,m) \dd m_1
\\
\nonumber
	& - \int_{m_+}^{M} \mathcal{K}(m_1,m) \dd m_1
\ ,
\end{align}
where we introduce $\mathcal{K}(m_1,m_2)=K(m_1,m_2)n(m_1)n(m_2)$ and combine the two integrals where possible. 

Clearly, if we introduce an additional kernel $\ceff(m_1,m_2)$ to force locality in the system we can ignore the two integrals in eq.~\eqref{eq:rewrite} that involve $m_0$ and $M$, and the collision integral is convergent. That is, when $\ceff\equiv 0$ if $m_1/m_2 < 1/q$ or $m_1/m_2 > q$ with $q>1$ finite, then eventually $m_0/m < 1/q$ and $M/m > q$ so taking the limits $m_0 \downarrow 0$ and $M \rightarrow \infty$ will not lead to divergence.

We leave $m_-$, $m$, $m_+$ fixed, and study the limit $m_0 \downarrow 0$. Because of the inequality~\eqref{eq:inequals}, we can use a Taylor expansion on the first integral in eq.~\eqref{eq:rewrite} and write
\begin{align}
\label{eq:lowend}
	&\approx -\int_{m_0}^{m_-} m_1 \partial_m\left(\mathcal{K}(m_1,m)\right) \dd m_1
\\
\nonumber
	&\approx -\partial_m (m^{\lambda-x}) \int_{m_0}^{m_-} m_1^{1-x} \dd m_1
\ .
\end{align}
Thus, for the integral~\eqref{eq:lowend} to converge as $m_0 \downarrow 0$, we require $x<2$.

Under the same conditions, we can write the last integral in eq.~\eqref{eq:rewrite} as
\begin{equation}
\label{eq:highend}
	\approx -m^{-x}\int_{m_+}^M m_1^{\lambda-x} \dd m_1
\ ,
\end{equation}
and if we let $M \rightarrow \infty$ we require $x > \lambda + 1$ for convergence of the integral. Both requirements $x<2$ and $x > \lambda + 1$ have been derived previously~\cite{crzab}.

Now, for differential sedimentation we have $\lambda = 4/3$ and $x = 13/6$, which implies divergence of the collision integral, and we require the additional kernel $\ceff$ with finite $q$ to force locality in this system so that we can achieve the scaling solution. For laminar shear we have $\lambda = 1$ and $x=2$, which puts it on the boundary of convergence.

The kernel of differential sedimentation is traditionally regarded as unphysical, because the growth rate of large particles increases faster than their actual volume~\cite{dongen}. It has been shown that by a simple modification of this kernel, scaling solutions can be realised in numerics with a particle system~\cite{horvai}. Both kernels, or at least their exponents $\lambda$ have also been discussed in relation to gelation, i.e. the formation of an infinite mass particle at finite time~\cite{crzab,dongen,dong1,dong2}. Within this family of kernels, the exponent $\lambda=1$ forms a boundary between gelling and non-gelling systems~\cite{djsmit,djsmit2}. The introduction of a finite large size cutoff limits the possibility of gelation in the IPS. However, the effect of a finite cutoff on the process of gelation in the SCE is unclear. 


\section{Differential sedimentation}
\label{sec:ds}

Numerical simulations of an interacting particle system have shown that a free merging system for differential sedimentation does not show the predicted scaling  eq.~\eqref{eq:ds}~\cite{horvai}. Instead, the particle density spectrum retrieved from the simulations can be explained by rewriting the SCE and assuming nonlocality of interactions to be the dominant process (see Appendix~\ref{app:nonlocal} and Ref.~\cite{horvai}). A similar approach was taken for the study of nonlocality in drift waves in turbulence~\cite{balk} and the evolution of galaxies~\cite{kontorovich}.

For differential sedimentation, we may write this approximate nonlocal solution to the SCE~\eqref{eq:smol} as
\begin{equation}
\label{eq:diffsednl}
	n(m)
\approx
	n_0 \left(\frac{m}{m_0}\right)^{-4/3} \exp \left[ W(\Theta) \left(\left( \frac{m}{m_0}\right)^{-1/3}-1 \right)\right]
\ ,
\end{equation}
with $n_0=n(m_0)$, and $W(\Theta)=W\left( 2 \frac{M}{m_0} \right)$ the Lambert W function or product log. Since we assume $M \gg m_0$, we may approximate $W(\Theta)$ by $\log\Theta$. 
\begin{figure}[htp]
\begin{center}
\includegraphics[width=0.7\textwidth]{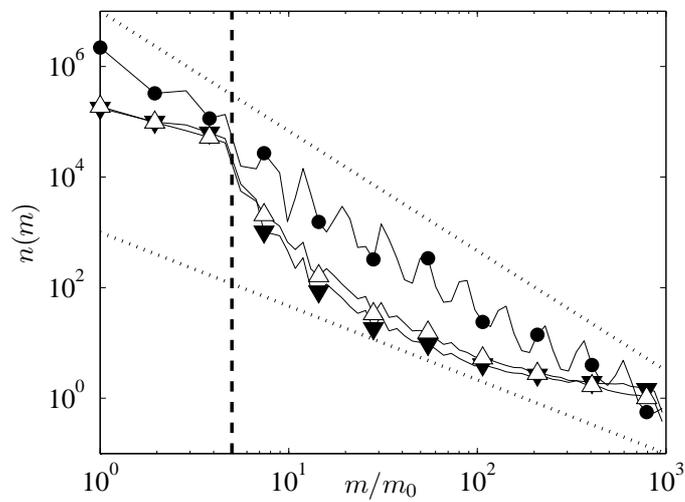}
\caption{Particle size density spectra for differential sedimentation. Circles: forced locality; upward triangles: free merging; downward triangles: forced nonlocality. Dotted lines show the $-13/6$ and $-4/3$ power laws. Dashed line signifies the end of the forcing range at $5 m_0$; cutoff for all simulations is at $M=10^3 m_0$. In all figures of mass spectrum, markers only indicate each graph, not all individual data points.}
\label{fig:ds}
\end{center}
\end{figure}
\begin{figure}[htp]
\begin{center}
\includegraphics[width=0.7\textwidth]{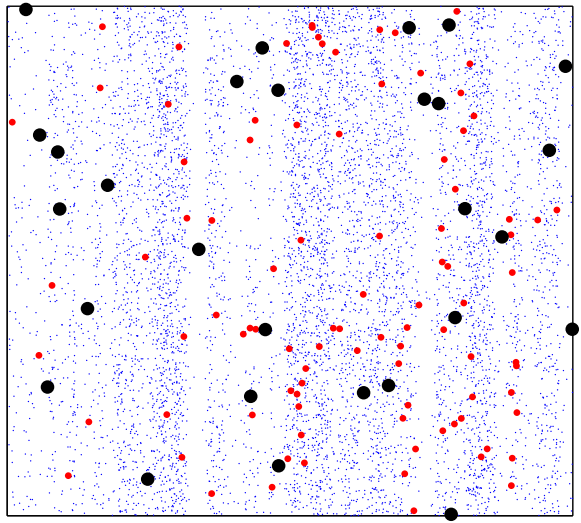}
\caption{Positions of all particles at a given time step in a vertical $(x,z)$-slice of width $0.5\mathrm{cm}$ of the domain for a free merging system with differential sedimentation and cutoff $M=10^4 m_0$. Small dots are particles of size $1-25 m_0$; intermediate are of size $25-625 m_0$; large are of size larger than $625 m_0$.}
\label{fig:shot}
\end{center}
\end{figure}
\begin{figure}
\begin{center}
\includegraphics[width=0.7\textwidth]{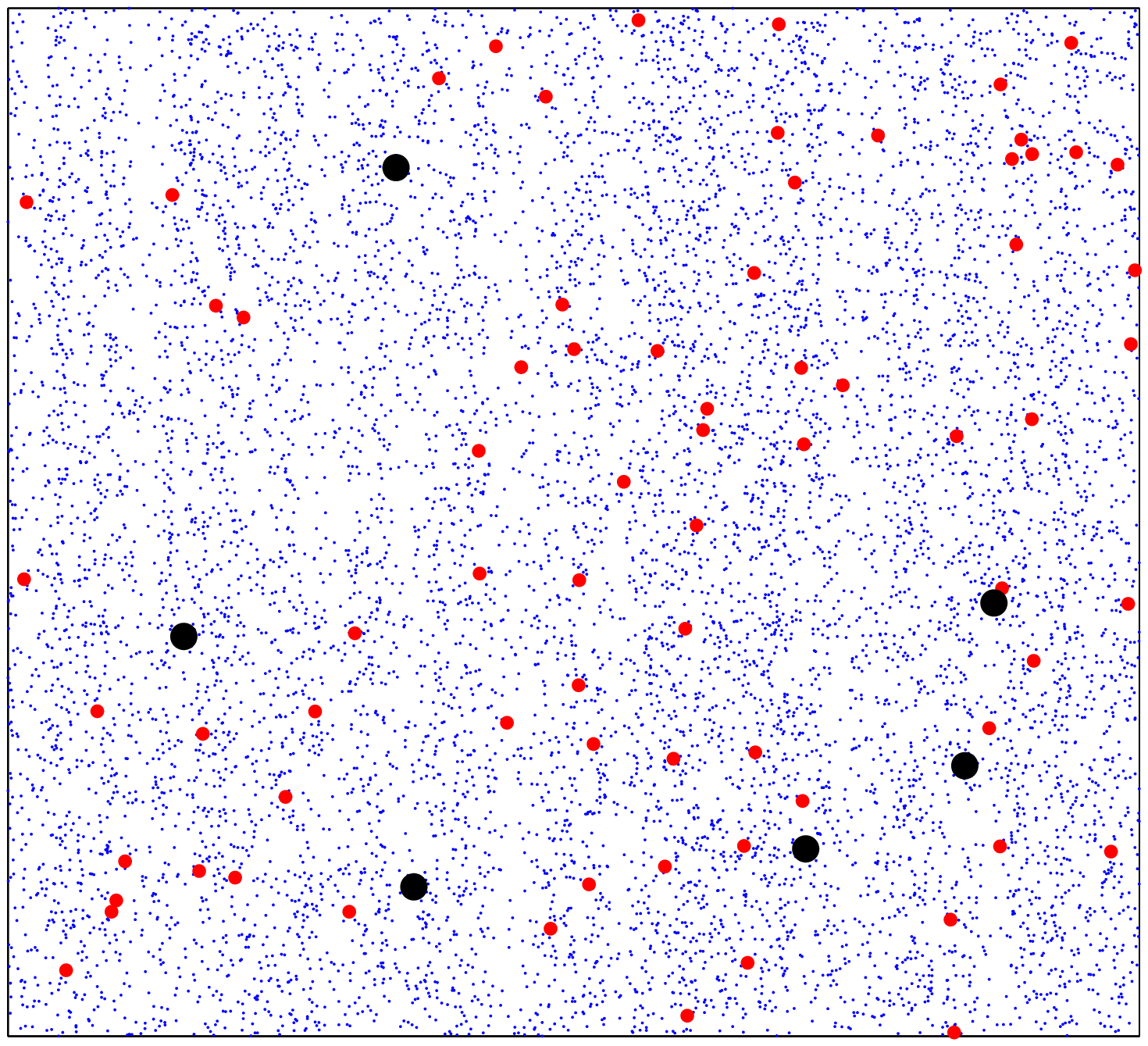}
\caption{Similar setup as Fig.~\ref{fig:shot} but with random reintroduction at $z=0$.}
\label{fig:shottoo}
\end{center}
\end{figure}

The dependence of the solution~\eqref{eq:diffsednl} on sharp cutoffs at $m_0$ and $M$ and the traditional solution of ``instantaneous gelation''~\cite{dongen} of similar kernels requires a word of caution concerning its validity as a steady state solution to the SCE. We will however check its applicability as an approximation to the IPS with cutoff. 

Note that at large $m$, the solution~\eqref{eq:diffsednl} follows the scaling $n(m)\sim m^{-4/3}$. This limit in itself is a conservation of number of particles for large $m$, i.e. there is a cascade for the number of large particles in the system. Since we do not use a strict monomer source $m_0$ we will not focus on a fit of eq.~\eqref{eq:diffsednl} towards smaller $m$.

We have performed numerical simulations for a system with differential sedimentation under different rules of collisions, as described in Sect.~\ref{sec:ips}. The data for forced locality in Fig.~\ref{fig:ds} shows rather strong fluctuations within the cascade, which we blame on the actual mechanism of differential sedimentation, in which particles of the same size will not merge. Thus, if we force particles of integer multiples $j$ of $m_0$ uniformly in the range $j\in[1,5]$, we have particles formed of size $3$ and $5$ due to coagulation in addition to the forcing, and we expect peaks in their distribution. Similarly, particles of size $4$ coagulate with sizes $2$, $3$, and $5$, but are not created (apart from forcing), and we expect a below average value for their number. These initial inhomogeneities are consequently carried through the cascade.

Also in Fig.~\ref{fig:ds} we plot the data for forced nonlocality and free merging. We clearly see that the free merging and nonlocal spectra do not follow a strict scaling law. Instead, we note that the simulation for free merging shows a spectrum similar to that of forced nonlocality, but neither can be convincingly fitted to a $-4/3$ power law at the large $m$ as predicted by the solution~\eqref{eq:diffsednl}. We see an overestimation of number of large particles closer to a spectrum $n(m)\sim m^{-1}$, which indicates a finite size effect or bottleneck in our simulation. 

\begin{figure}[htp]
\begin{center}
\includegraphics[width=0.7\textwidth]{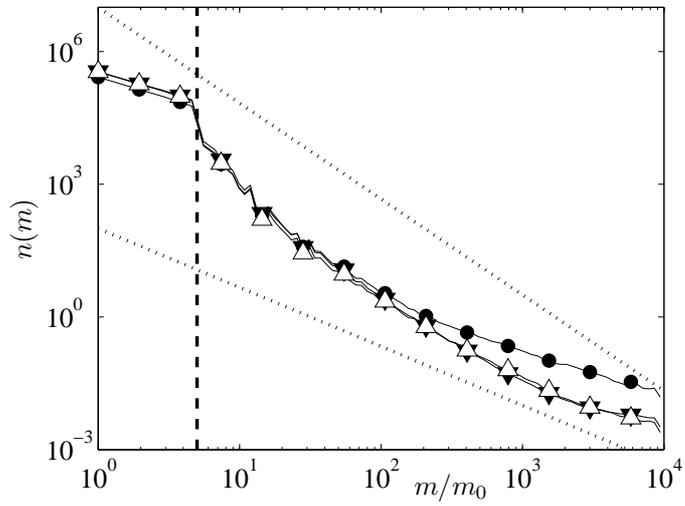}
\caption{Free merging simulations with cutoff $M=10^4 m_0$. Circles: cubic domain; upward triangles: cubic with random reintroduction; downward triangles: no random reintroduction, but domain size $5\times 5\times 40$. Dotted lines show the $-13/6$ and $-4/3$ power laws. Dashed line signifies the end of the forcing range at $5 m_0$.}
\label{fig:dstoo}
\end{center}
\end{figure}

A snapshot of the free merging system at large times shows that large particles have dug out channels void of any particles (Fig.~\ref{fig:shot}). Due to periodicity of the system, the large particle will have traversed the whole length in a time too short for smaller sized particles to hang around and not be merged with the larger particle. This causes stagnation of the mass flux and consequently a bottleneck accumulation on particle mass distribution. As we see, such a bottleneck mechanism has a non-mean field origin. We try to avoid this behaviour, which contradicts the mean field assumption used in the derivation of the SCE, by stretching the domain in the direction of motion. The result for such a system is shown in Fig.~\ref{fig:dstoo}. We note that the bottleneck effect is not as strong as in the cubic domain, and the large particle behaviour is close to the nonlocal solution~\eqref{eq:diffsednl}.

Independently, we may try to avoid the non-mean field behaviour by adapting the periodicity of the system, effectively introducing mixing at a particular level of $z$. When a particle leaves the system at $L_z +z'$, where $L_z$ is the length of the domain in the direction of motion, we reintroduce the particle at $z'$, but with new and random $x,y$-coordinates. The resulting spectrum for this system is also shown in Fig.~\ref{fig:dstoo}, and a snapshot is given in Fig.~\ref{fig:shottoo}. We note that the channels for large particles are not throughout the system anymore, and that there are typically fewer large particles. Interestingly, the spectrum is not significantly different from the spectrum for the stretched domain. A similar modification of periodicity has been used previously for an IPS with laminar shear~\cite{VLP1}.
\begin{figure}[htp]
\begin{center}
\includegraphics[width=0.7\textwidth]{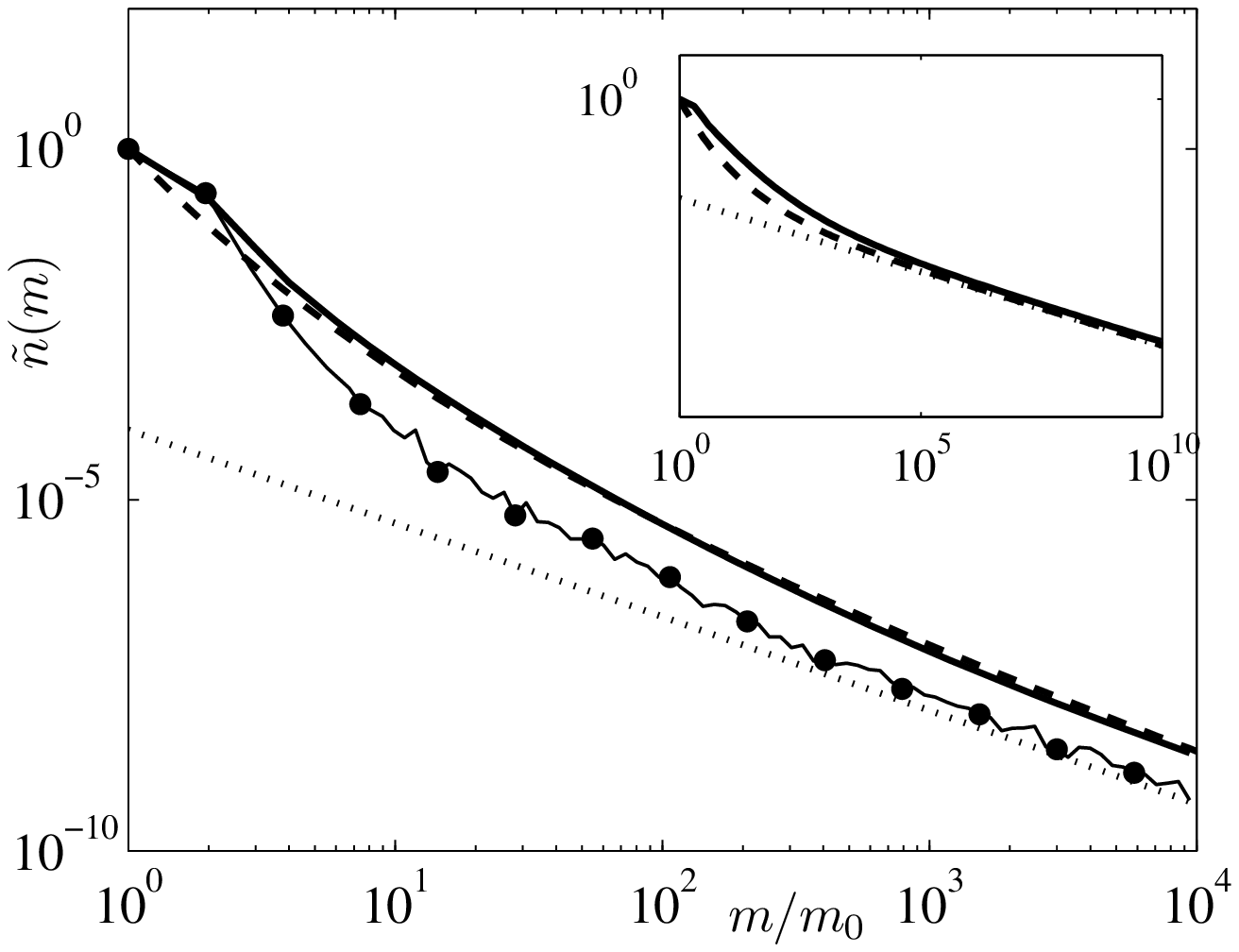}
\caption{Free merging simulations with increasing cutoff $M$. $M=10^4 m_0$ in cubic domain and with random reintroduction (circles). Solid line: SCE numerics with $M=10^4 m_0$. Dashed line: nonlocal solution~\eqref{eq:diffsednl}. Inset: SCE numerics and nonlocal solution for $M=10^{10} m_0$. Spectra shown are for $\tilde{n}(m)=n(m)/n(m_0)$. Dotted line shows the $-4/3$ power law. SCE numerics courtesy Colm Connaughton.}
\label{fig:five}
\end{center}
\end{figure}

In an attempt to gain a longer range of $n(m)\sim m^{-4/3}$ for large $m$, we compare the data for the IPS with data from direct numerical simulations of the SCE with cutoff~\eqref{eq:cuts}. The results from Fig.~\ref{fig:five} have been realised with forcing of equal volume into both $m_0$ and $2 m_0$. A wider inertial range has been realised with the SCE numerics, and we see the same qualitative behaviour of the spectrum with $-4/3$ at large $m$ for all simulations.

From Fig.~\ref{fig:five} we see good agreement between the analytic solution~\eqref{eq:diffsednl} and the SCE numerics. This is not entirely surprising, since this solution was derived from the SCE with cutoff, but using the assumption of dominance of nonlocal interactions. Both the SCE numerics and the analytic solution overestimate the IPS of which they are an approximation. The deviation of the SCE from the IPS may be due to non-mean field behaviour and the finite domain, or it could be related to gelation. The exact nature of this deviation will require further study of the IPS and the influence of the source and cutoff.


\section{Laminar shear}
\label{sec:ls}

For laminar shear we cannot find a nonlocal solution of the form~\eqref{eq:diffsednl} (see Appendix~\ref{app:nonlocal}). Instead, we present the results from the IPS and look for differences between results for the different rules of collision.

\begin{figure}[htp]
\begin{center}
\includegraphics[width=0.8\textwidth]{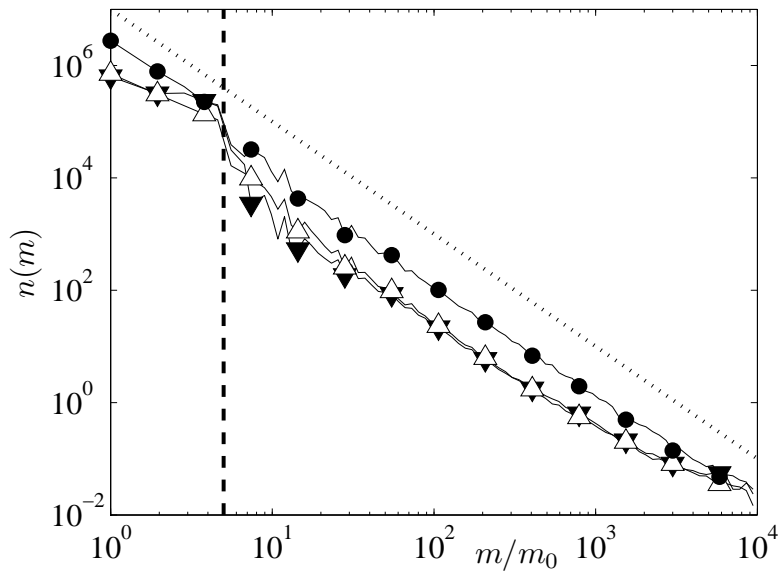}
\caption{Particle size density spectra for laminar shear. Circles: forced locality; upward triangles: free merging; downward triangles: forced nonlocality. Dotted line shows the $-2$ scaling solution. Dashed line signifies the end of the forcing range at $5 m_0$; cutoff for all simulations is at $M=10^4 m_0$.}
\label{fig:shear}
\end{center}
\end{figure}

The particle spectra for all three different setups are shown in Fig.~\ref{fig:shear}, and we note that the simulation for forced locality clearly follows the scaling solution $n(m)\sim m^{-2}$. Also, the fluctuations down the spectrum that were so clear for differential sedimentation are not apparent here, so we assume that these were indeed due to the specific kernel.
\begin{figure}[htp]
\begin{center}
\includegraphics[width=0.8\textwidth]{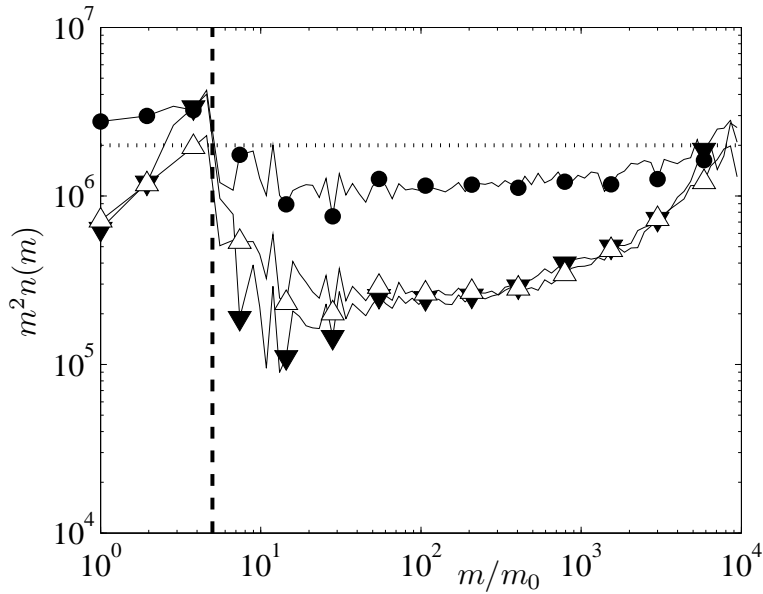}
\caption{Compensated particle size density spectra, $m^2 n(m)$ for the same setup as Fig.~\ref{fig:shear}.}
\label{fig:shearlogs}
\end{center}
\end{figure}

Again, the free merging system is similar to the nonlocal system in the particle spectrum, and we note that near the source and sink they deviate from the $-2$ scaling. The absence of this deviation in the data for forced locality clearly indicates that this is a result of nonlocality. In Fig.~\ref{fig:shearlogs} we show the compensated spectra $m^2 n(m)$ for all simulations. We see that the deviation from the $-2$ power law for free merging and forced nonlocality is a slowly evolving function of mass.

The results from Fig.~\ref{fig:shearlogs} combined with the knowledge that the exponent $\lambda=1$ for the SCE is critical in terms of convergence and gelation leads us to believe we should look for corrections to the $-2$ spectrum. Following Kraichnan's work in two-dimensional turbulence~\cite{kraichnan1,kraichnan2}, where he predicted logarithmic corrections for the enstrophy cascade which also has marginal locality, we assume the corrections to the $-2$ scaling solutions to be logarithmic as well.


\section{Discussion}
\label{sec:discuss}

We have performed numerical simulations of an interacting particle system (IPS) under different sets of rules for both differential sedimentation and laminar shear. The system does not assume mean field behaviour, yet provides results that are remarkably consistent with theoretical predictions using the SCE for the respective mechanisms of coagulation.

The IPS with forced locality confirms the KZ scaling laws for the SCE, whilst a free merging system deviates from said solutions, and its spectrum is more similar to a system that assumes nonlocal interactions to be dominant. For differential sedimentation we have predicted a solution which has a particle number cascade of $-4/3$ at large $m$. Our numerics show spectra close to this behaviour, but the simulations suffer from a finite size effect. In the standard system, due to periodicity large particles dig out channels which become devoid of smaller particles, which slows down the mass cascade. This in turn leads to a bottleneck accumulation of mass at large $m$, which behaviour is non-mean field in origin. We limit this effect by stretching the system or by introducing mixing, both of which numerics show a $-4/3$ spectrum for large $m$.

The exponent $\lambda=1$ in the kernel for laminar shear is known to be marginal in terms of convergence of the collision integral~\cite{crzab} and we recognise this marginal nonlocal behaviour in the results from our IPS. In a compensated spectrum we note that corrections to the $-2$ scaling are necessary, and we will investigate such corrections in a future paper.

Traditionally, both differential sedimentation and laminar shear are mechanism thought of as leading to ``instantaneous gelation''~\cite{dongen,dong1}. We note that in spite of this prediction of unphysical behaviour, we find a persistent solution with our IPS, which can be described by an approximate solution using the SCE with cutoffs.

Finally, we note that the remarkably good agreement of the IPS with the predictions for the SCE with cutoffs has also been confirmed by simulations done for self-similarity~\cite{horvai}, and the nonlocal prediction of $n(m)\sim m^{-4/3}$ for large mass can also be recognised in SCE numerics.

\section*{Acknowledgements}
\label{sec:acknowledge}

We would like to thank Colm Connaughton and Oleg Zaboronski for helpful discussions and suggestions. We also thank Colm for providing us with the SCE numerics.


\appendix
\setcounter{section}{0}

\section{Nonlocal solutions to the collision integral}
\label{app:nonlocal}

We continue the analysis in Sec.~\ref{sec:convergence} from eq.~\eqref{eq:rewrite}, forgetting the assumption of a scaling solution for $n(m)$, and instead assume that nonlocal interactions are dominant in the system. Again leaving $m_-$, $m$, $m_+$ fixed, we can use eq.~\eqref{eq:lowend} and eq.~\eqref{eq:highend} to write 
\begin{equation}
\label{eq:nonlocpde}
	\frac{D n}{D t}
\approx
	-c_1 \partial_m (m^{\lambda} n(m)) - c_2 n(m)
\ ,
\end{equation}
where we replaced $m^{-x}$ by $n(m)$, and we have replaced the integrals in eq.~\eqref{eq:lowend} and~\eqref{eq:highend} by $c_1$ and $c_2$ respectively, again replacing $m^{-x}$ by $n(m)$.

Now, we still assume that the inequalities~\eqref{eq:inequals} hold, but we keep $m_0$ and $M$ fixed. Assuming this system may reach steady state, we obtain an ODE for eq.~\eqref{eq:nonlocpde} which for $\lambda > 1$ has the solution
\begin{equation}
\label{eq:nonlcsoln}
	n(m)
=
	C m^{-\lambda} \exp \left[\frac{1}{\lambda-1}\frac{c_2}{c_1} m^{1-\lambda}\right]
\ .
\end{equation}
We will not use this approach to find a solution for the marginal case with $\lambda=1$.

If we define $n_0=n(m_0)$, and set $m=m_0$ in eq.~\eqref{eq:nonlcsoln}, we may write
\begin{equation}
\label{eq:C}
	C
=
	n_0 m_0^{\lambda} \exp \left[ \frac{1}{1-\lambda} \frac{c_2}{c_1} m_0^{1-\lambda} \right]
\ ,
\end{equation}
and we write for the solution~\eqref{eq:nonlcsoln}~\footnote{Alternatively, one can match the constant $C$ to the flux of particles near $m_0$.}
\begin{equation}
\label{eq:n-C}
	n(m)
=
	n_0 \left( \frac{m_0}{m} \right)^{\lambda} \exp \left[ \frac{1}{\lambda-1} \frac{c_2}{c_1} (m^{1-\lambda} - m_0^{1-\lambda}) \right]
\ .
\end{equation}
We may also find approximations for $c_2$ and $c_1$ in order to find a steady state solution
for $\lambda>1$. Using eq.~\eqref{eq:n-C} we write
\begin{equation}
\label{eq:c-2a}
	c_2
=
	\int^M n_0 m_0^{\lambda} \exp \left[ A (m_1^{1-\lambda}-m_0^{1-\lambda}) \right] \dd m_1
\ ,
\end{equation}
with $A=\frac{1}{\lambda-1} \frac{c_2}{c_1}$. Since we have $M,m \gg m_0$, and
$1-\lambda < 0$, we may write
\begin{equation}
\label{eq:c-2done}
	c_2
\approx
	n_0 m_0^{\lambda} M \exp \left[ -A m_0^{1-\lambda} \right]
\ .
\end{equation}
For $c_1$ we write
\begin{equation}
\label{eq:c-1a}
	c_1
=
	\int_{m_0} n_0 m_0^{\lambda} m_1^{1-\lambda} \exp \left[ A (m_1^{1-\lambda}-m_0^{1-\lambda}) \right] \dd m_1
\ ,
\end{equation}
and we note that as $m_1 \to m_0$, the term $\exp[...]\to 1$. Thus, we may write
\begin{equation}
\label{eq:c-1done}
	c_1
\approx
	n_0 m_0^{\lambda} \left[ \frac{1}{2-\lambda} m^{2-\lambda} \right]_{m_0}
\approx
	\frac{1}{2-\lambda} n_0 m_0^2
\ ,
\end{equation}
where we note the additional restriction that $\lambda < 2$.

Now we can write
\begin{equation}
\label{eq:exponent}
	\frac{1}{\lambda-1}\frac{c_2}{c_1}
=
	A
=
	\frac{2-\lambda}{\lambda-1} m_0^{\lambda-2} M \exp \left[ -A m_0^{1-\lambda} \right]
\ .
\end{equation}
Let us write $A= \theta m_0^{\lambda-1}$, and find
\begin{equation}
\label{eq:a}
	\theta
=
	\frac{2-\lambda}{\lambda-1} \frac{M}{m_0} e^{-\theta}
\ ,
\end{equation}
or
\begin{equation}
\label{eq:a2}
	\theta e^{\theta}
=
	\frac{2-\lambda}{\lambda-1} \frac{M}{m_0}
\ ,
\end{equation}
which has as a solution for $\theta$
\begin{equation}
\label{eq:lambert}
	\theta
=
	W\left( \frac{2-\lambda}{\lambda-1} \frac{M}{m_0} \right)
=
	W(\Theta)
\ ,
\end{equation}
where $W(z)$ is the Lambert W function or product log. This is a single-valued, real function under our assumption $1 < \lambda < 2$. 

We find our general solution to the Smoluchowski equation with nonlocality
and homogeneity $\lambda>1$ to be
\begin{equation}
\label{eq:solution}
	n(m)
\approx
	n_0 m_0^{\lambda} m^{-\lambda} \exp \left[ W(\Theta) \left(\left( \frac{m}{m_0}\right)^{1-\lambda}-1 \right)\right]
\ .
\end{equation}
Note that such a solution should be taken with caution, as it uses the assumption of sharp cutoffs $m_0$ and $M$. As such, the exponential ``spike'' towards small $m$ will depend on the forcing near $m_0$. We do expect the large $m$ scaling of $n(m)\sim m^{-\lambda}$ to be a verifiable feature of the nonlocal solution.


\bibliographystyle{unsrt}
\bibliography{dissertation}


\end{document}